\documentclass[aip,rsi,reprint,graphicx]{revtex4-1} 
\usepackage{graphicx}

\draft 

\begin{document}


\title{Note: Cryogenic microstripline-on-Kapton microwave interconnects}

\author{A.I.~Harris}
 \email{harris@astro.umd.edu}
 \affiliation{Department of Astronomy, University of Maryland,
   College Park, MD 20742.}

\author{M.~Sieth}
\author{J.M.~Lau}
\author{S.E.~Church}
\affiliation{Department of Physics, Stanford University, Palo Alto, CA
94305}

\author{L.A.~Samoska}
\affiliation{Jet Propulsion Laboratory, MS~168-3, California Institute of
  Technology, Pasadena, CA 91109.}

\author{K.~Cleary}
\affiliation{Department of Astronomy, MS~222, California Institute of
  Technology, Pasadena, CA 91125.}

\date{\today}

\begin{abstract}
  Simple broadband microwave interconnects are needed for increasing
  the size of focal plane heterodyne radiometer arrays.  We have
  measured loss and cross-talk for arrays of microstrip transmission
  lines in flex circuit technology at 297 and 77\,K, finding good
  performance to at least 20\,GHz.  The dielectric constant of Kapton
  substrates changes very little from 297 to 77\,K, and the electrical
  loss drops.  The small cross-sectional area of metal in a printed
  circuit structure yields overall thermal conductivities similar to
  stainless steel coaxial cable.  Operationally, the main performance
  tradeoffs are between crosstalk and thermal conductivity.  We tested
  a patterned ground plane to reduce heat flux.
\end{abstract}


\maketitle

\section{Introduction}

Simple broadband microwave interconnects are key components for arrays
of focal plane heterodyne radiometers.  Here we report on an
investigation of transmission line arrays printed on flexible circuit
board substrates, part of a program to develop and assess components
for a scalable millimeter-wave focal plane radiometer\cite{sieth12}.
For tens of focal plane elements, microwave intermediate frequency
signals can be routed on individual semi-rigid coaxial cables, but
this approach becomes cumbersome for large focal plane arrays.  Here
we report on an alternative interconnect: microstripline on a
polyimide (Kapton\cite{kaptonTM}) flex circuit substrate.
A number of papers\cite{oliver09, oliver10, mcgibney11}, among others,
report microwave characterization of flex substrates, but none that we
are aware of report cryogenic properties.

Of the well-developed planar transmission line structures, we choose
microstripline because of its mechanical simplicity, its relatively
low electrical loss, and because it requires the least metal of common
planar transmission lines.  The last item is an important
consideration for transmission lines between components at different
temperatures.  In spite of copper's high thermal conductivity, the
small metallic cross-section of planar lines results in a total heat
flow along a flex circuit comparable to that through the much larger
cross-sections of stainless steel and steel in standard cryogenic
coaxial cable.  With strength carried by the Kapton substrate, and
high-frequency fields confined to a thin layer near the conductor
surfaces by the skin effect, very thin conductors are practical for
the lines.  The thinnest standard copper cladding on Kapton is
0.5\,oz.\ per square foot, or 0.7\,mil (0.0007\,in, $18\,\mu$m) thick.
Calculated microstrip/coax heat flow ratios for typical thermal
conductivities\cite{scott59} along an 8-circuit evaluation structure
(8 parallel 11 mil strips on an 850 mil wide ground plane) divided by
that along 8 stainless-steel 085 coaxial cables are 0.66 for
297\,K--77\,K end temperatures, and 2.9 for 77\,K--20\,K.  For
microstriplines most of the thermal path is in the ground plane, so we
investigated a patterned ground plane that reduced the amount of
metal.  

We report measurements on two test structures.  The first included
microwave resonators to evaluate materials properties at 297 and 77\,K
(room temperature and liquid nitrogen).  The second was a set of 8
parallel microstrip lines to evaluate multiple-line performance over
two ground plane patterns.  Both structures were on DuPont Pyralux
AP-8555R stock, which has 0.5\,oz/ft$^2$ rolled\cite{oliver09} copper
bonded to both sides of 5\,mil thick polyimide substrate material.
Although thinner substrates are available, we chose 5\,mil mostly
because of fabrication tolerances: with this thickness, the microstrip
lines are 11\,mils wide, and a 10\% width fabrication error still
produces a line with an impedance close to $50\,\Omega$.  As secondary
considerations, electrical loss drops with increasing strip width, and
an 11\,mil width is suitable for solder contact between SMA connector
pins and the strip.  Most flex circuits have a thin dielectric
coverlay to protect the traces, but the adhesive has high electrical
loss at microwave frequencies\cite{oliver09}, so the test structures
had no coverlays. 

\section{Tee resonator measurements at 297 and 77\,K}

A commercial firm fabricated transmission line test structures on a
1.5\,inch by 4.1\,in substrate with a full ground plane on one side
and three microstrip transmission lines across the short dimension on
the other side.  Two of the lines had perpendicular shunt open stubs
to make tee resonators, with stub lengths 3.000 and 0.550\,inch long.
Johnson 142-0701-851 edge connectors with 10\,mil diameter pins made
contact with the lines and ground pads to either side of the line.
Gold plating (5\,$\mu$in of gold on 100\,$\mu$in nickel) kept the
traces from oxidizing while allowing wire bond and solder connections.

The substrate shrinks little between room temperature and 77\,K while
remaining flexible.  There was no apparent bending or other thermally
induced stress between the substrate and fully metalized ground plane
when the test structure was immersed in liquid nitrogen. We corrected
for resonator length change with cooling by measuring the length of
the test substrate at room temperature and when immersed with a
stainless steel scale in a shallow tray filled with liquid nitrogen;
the length change on cooling was only 5\,mil on the scale. After
correcting for the scale's own fractional length contraction from
293--77\,K\cite{nistSS} we derived a substrate fractional length
contraction of $\Delta L/L = -4.0\times10^{-3}$ between 297 and 77\,K.
At 77\,K, the 3\,inch resonator was shorter by 13\,mils.

All resonator measurements were in vacuum with the substrate attached
to the cold plate of a small liquid nitrogen cryostat. A copper
radiation shield attached to the cold plate and lined with microwave
absorber covered the substrate to block infrared radiation that would
otherwise heat the substrate.  Comparison of warm transmission with
and without the copper cover showed that the cover did not affect
microwave transmission.  Conformable 085 coaxial cables, 8\,in long,
connected the test structure to hermetic SMA feedthroughs passing
through the cryostat wall; 3\,dB attenuators between the cables and
feedthroughs helped reduce residual standing waves. We made
measurements of the resonators and the through on the test structure
at 297 and 77\,K with an Agilent 8722D vector network analyzer,
sampling 201 points from 50\,MHz to 20\,GHz.

\begin{figure}[t]
\includegraphics[width=3in, bb=81 290 525 583, clip=TRUE]{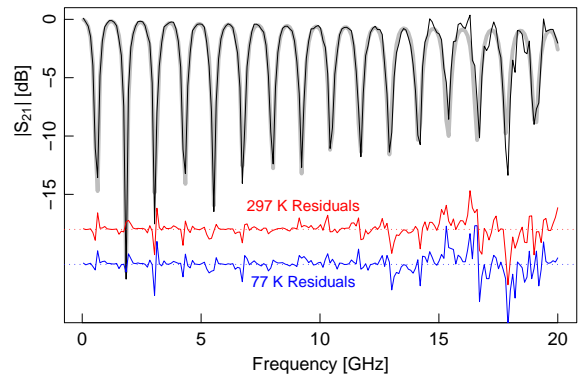}
\caption{Comparison of measured data (thin black line) and model fit
  from 0.05 to 12\,GHz (grey line) for the cold resonator.  Curves
  below show residuals from the fit for warm and cold resonators on
  the same dB scale but with offsets indicated by the dotted lines.
  This plot shows that the fit is excellent to about 13.5\,GHz and is
  representative to at least 20\,GHz.  \label{fig:fitPlot}}
\end{figure}

Dividing the test structure's resonator transmission ($|S_{21}|$) by
that of the through line removed cable, attenuator, and connector
losses, giving a clean measurement of the tee resonator alone.  We
used Microwave Office\cite{MWO} (MWO) for this division and to fit for
substrate dielectric constant and loss tangent by comparing the
derived values of $|S_{21}|$ to those from MWO's parametric models
(which include finite-element electromagnetic calculations for the
discontinuities at the tee) and optimization function.  The topmost
lines in Figure~\ref{fig:fitPlot} display the cold measurement and the
MWO model fit for the 3\,in resonator.  Plots of the residuals between
the two, as well as the residual for the warm measurements, shows that
the models are good representations at both temperatures.  Structure
at higher frequencies is common to both warm and cold resonators and
is probably due to different connector mismatches in the resonator and
through lines.  The overall agreement between measurement and theory
indicates that the materials parameters we derive are valid to at
least 20\,GHz.

Table~\ref{tab:table} summarizes derived electrical properties.  The
dielectric constant $\epsilon_r$ changed by a negligible amount
between room temperature and 77\,K; a representative value for
microwave frequencies is $\epsilon_r = 3.37$.  There is a weak
dependence on the fit frequency range, 0.05\,GHz to $f_{\rm upper}$,
for $\epsilon_r$, which is just visible in plots for different
frequencies.  Both dielectric and metalization losses contribute to
overall loss, but the fits were insensitive to losses in the metal,
parametrized by conductivity $\rho$ relative to gold, and we used a
value of $\rho = 0.7$.  Loss in the dielectric drops by a factor of
approximately two on cooling, from $\tan \delta = 0.013$ to 0.007.

\begin{table}
  \caption{Dielectric constants $\epsilon_r$ and loss tangents
    $\tan \delta$ derived from resonator measurements at 297 and 77\,K,
    with fitting from 0.05\,GHz to $f_{\rm upper}$.  The change in
    $\epsilon_r$ with frequency shows a small amount of
    frequency dependence.  The change in
    $\epsilon_r$ with temperature is negligible for most purposes. Loss has a clear
    temperature dependence, however, with $\tan \delta$ dropping
    by a factor of $\sim 2$ on cooling.  \label{tab:table}}
\begin{ruledtabular}
\begin{tabular}{rrrrr}
$f_{\rm upper} [$GHz$]$ &
$\epsilon_{r, {\rm 297\,K}}$ & $\epsilon_{r, {\rm 77\,K}}$ &
$\tan \delta_{, \rm 297\,K}$ & $\tan \delta_{, \rm 77\,K}$ \\
\hline
3 & 3.378 & 3.377 & 0.008 & 0.000 \\
6 & 3.372 & 3.370 & 0.012 & 0.006 \\
12 & 3.348 & 3.350 & 0.013 & 0.007 \\
\end{tabular}
\end{ruledtabular}
\end{table}

\section{Loss and cross-talk with parallel lines at 297\,K}

\begin{figure}[t]
\includegraphics[width=3in]{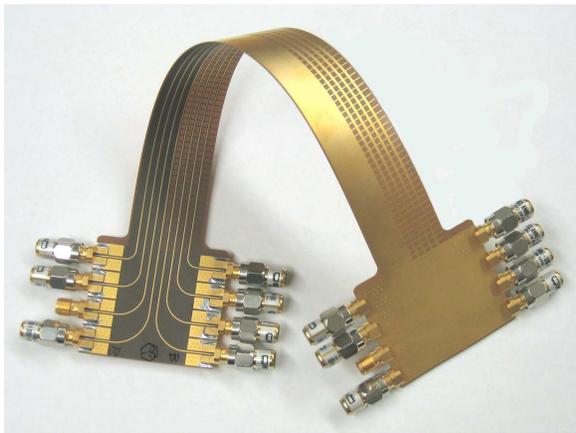}
\caption{Photograph of parallel line test structure, with a twist to
  show both the transmission line
  layout (left side) and the ground plane patterning (right side).
\label{fig:parallellines}}
\end{figure}

We directly assessed the performance of microwave interconnects with a
structure with eight parallel microstrip transmission lines with strip
widths of 11\,mil on 100\,mil centers (gaps between lines equal to 8.1
times the strip widths).  The minimum line length was 8.5\,in, and the
maximum was 11.7\,in.  As shown in Figure~\ref{fig:parallellines},
four of the lines were over a solid ground plane, and the other four
were over a patterned ground plane with reduced thermal conductivity.
The patterning was a solid plane 55\,mils wide below each 11\,mil-wide
line, providing termination for most of the field lines, with
10\,mil-wide cross-strips on 80\,mil centers tying the grounds
together across the width of the structure.  The cross-strips must
have spacing with distance well below a quarter of the shortest
wavelength to avoid resonances between the lines, and $\lambda/10$ or
closer to reduce structure in $S_{21}$.  With a pattern of relatively
narrow ground planes under the lines, joined by thin cross-connects,
the calculated heat flow along the structure is a factor of three
lower than a solid ground plane.  Broader ground strips under the
lines would reduce cross talk at the cost of higher thermal
conductivity.  Reducing line widths on a thinner substrate is an
additional attractive solution at low temperatures, where the
electrical loss is lower, although fabrication tolerances may become
critical.

\begin{figure}[h]
\includegraphics[width=3in, bb=81 290 525 593, clip=TRUE]{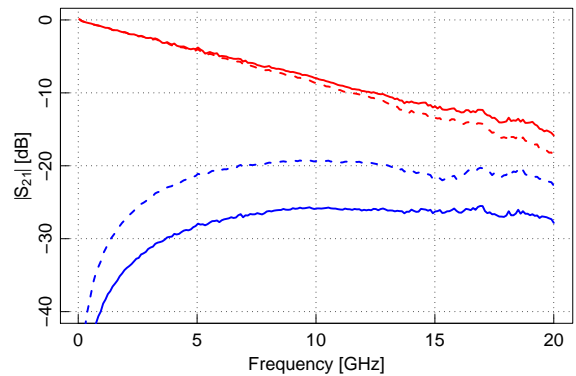}

\caption{Transmission and nearest-neighbor cross-coupling for solid
  (solid lines) and patterned (dashed lines) ground planes for a
  10.6\,inch line with 0.1\,inch spacing to its neighbors.
  Transmission (upper pair of lines) is only slightly affected by the
  missing metal in the ground plane, but cross-coupling (lower pair of
  lines) is sensitive to patterning.
  \label{fig:compare}}
\end{figure}

Figure~\ref{fig:compare} shows the transmission loss and
cross-coupling for the 10.6\,inch lines over the two ground planes.
Cross-coupling in this plot is to a nearest-neighbor line with
8.2\,inches of parallel run; the nearest neighbor on the other side
has 8.8\,inches of parallel run.  The transmission loss is only
slightly higher for the patterned ground plane, but the cross talk is
substantially higher: the far-out field lines carry little power but
are responsible for cross-coupling, and are poorly terminated on the
strips between the transmission lines.

Transmission loss for this structure with a solid ground plane is
closely 0.076\,dB/GHz/in. A model fit over 0.05\,GHz--12\,GHz gives
$\tan \delta = 0.018$, slightly higher than the fit to the
room-temperature resonator data, $\tan \delta = 0.013$.  Loss from
cross-coupling between lines is present and is accurately predicted by
coupled-line microstrip theory.  Modeling for other spacings shows
cross-talk decreases with increasing spacing and frequency, but has
periodic maxima with line length, as expected for a forward-coupled
pair of lines\cite{ikalainen87}: MWO calculations yield 20\,dB maxima
for gaps of 6.5 time the strip widths, and 30\,dB for 12 times the
strip widths.  For comparison, standard 085 semi-rigid cryogenic
coaxial cable (stainless steel outer jacket, Teflon insulation,
silver-plated steel wire center conductor) has a loss of about
0.0125\,dB/GHz/in and essentially infinite isolation, but with little
possibility for mass connection.

\begin{acknowledgments}
  This work was supported by NSF grant ATI-0905855 (ARRA).  We thank
  AWR for providing access to Microwave Office under its University
  Program.  We benefited from conversations with Dr.\ M.~Morgan of the
  National Radio Astronomy Observatory.
\end{acknowledgments}


%

\end{document}